\documentclass[%
 aip,
 amsmath,amssymb,
 reprint,%
]{revtex4-1}
\usepackage{color}
\usepackage{graphicx}
\usepackage{dcolumn}
\usepackage{bm}
\usepackage{textcomp}
\usepackage[utf8]{inputenc}

\usepackage[utf8]{inputenc}
\usepackage[T1]{fontenc}
\usepackage{mathptmx}
\usepackage{etoolbox}

\makeatletter
\def\@email#1#2{%
 \endgroup
 \patchcmd{\titleblock@produce}
  {\frontmatter@RRAPformat}
  {\frontmatter@RRAPformat{\produce@RRAP{*#1\href{mailto:#2}{#2}}}\frontmatter@RRAPformat}
  {}{}
}%
\makeatother
\begin{document}

\preprint{AIP/123-QED}

\title{Single-shot laser-pulse-induced magnetization reversal in CoFeB/MgO-based magnetic tunnel junctions}
\author{Junta Igarashi} 
\email{junta.igarashi@aist.go.jp}
\affiliation{Universit\'{e} de Lorraine, CNRS, IJL, Nancy, France}
\affiliation{Global Research and Development Center for Business by Quantum-AI technology (G-QuAT), National Institute of Advanced Industrial Science and Technology, Tsukuba, Japan}
\author{S\'{e}bastien Geiskopf}
\affiliation{Universit\'{e} de Lorraine, CNRS, IJL, Nancy, France}
\author{Takanobu Shinoda}
\affiliation{Laboratory for Nanoelectronics and Spintronics, Research Institute of Electrical Communication, Tohoku University, Sendai, Japan}
\affiliation{Department of Electronic Engineering, Tohoku University, Sendai, Japan}
\author{Butsurin Jinnai}
\affiliation{Advanced Institute for Materials Research, Tohoku University, Sendai, Japan}
\author{Yann Le Guen}
\affiliation{Universit\'{e} de Lorraine, CNRS, IJL, Nancy, France}

\author{Julius Hohlfeld}
\affiliation{Universit\'{e} de Lorraine, CNRS, IJL, Nancy, France}
\author{Shunsuke Fukami}
\affiliation{Laboratory for Nanoelectronics and Spintronics, Research Institute of Electrical Communication, Tohoku University, Sendai, Japan}
\affiliation{Department of Electronic Engineering, Tohoku University, Sendai, Japan}
\affiliation{Advanced Institute for Materials Research, Tohoku University, Sendai, Japan}
\affiliation{Center for Science and Innovation in Spintronics, Tohoku University, Sendai, Japan.}
\affiliation{Center for Innovative Integrated Electronic Systems, Tohoku University, Sendai, Japan.}
\affiliation{Universit\'{e} de Lorraine, CNRS, IJL, Nancy, France}
\affiliation{Inamori Research Institute for Science, Kyoto, Japan.}

\author{Hideo Ohno}
\affiliation{Laboratory for Nanoelectronics and Spintronics, Research Institute of Electrical Communication, Tohoku University, Sendai, Japan}
\affiliation{Department of Electronic Engineering, Tohoku University, Sendai, Japan}
\affiliation{Advanced Institute for Materials Research, Tohoku University, Sendai, Japan}
\affiliation{Center for Science and Innovation in Spintronics, Tohoku University, Sendai, Japan.}
\affiliation{Center for Innovative Integrated Electronic Systems, Tohoku University, Sendai, Japan.}
\author{Jon Gorchon}
\affiliation{Universit\'{e} de Lorraine, CNRS, IJL, Nancy, France}
\author{St\'{e}phane Mangin}
\affiliation{Universit\'{e} de Lorraine, CNRS, IJL, Nancy, France}
\affiliation{Center for Science and Innovation in Spintronics, Tohoku University, Sendai, Japan.}
\author{Michel Hehn}
\affiliation{Universit\'{e} de Lorraine, CNRS, IJL, Nancy, France}
\affiliation{Center for Science and Innovation in Spintronics, Tohoku University, Sendai, Japan.}
\author{Gr\'{e}gory Malinowski}
\affiliation{Universit\'{e} de Lorraine, CNRS, IJL, Nancy, France}

\date{\today}

\begin{abstract}
We demonstrate single-shot laser-pulse-induced magnetization reversal in rare-earth-free CoFeB/MgO magnetic tunnel junctions (MTJs), a material system widely adopted in spin-transfer torque magnetic random-access memory (STT-MRAM). By tuning the Ru capping layer thickness, we modify the laser energy absorption profile and observe magnetization reversal from the parallel (P) to antiparallel (AP) state, with switching observed for $t_\text{Ru} \geq 2.0\,$ nm. Furthermore, we detect magnetization reversal in a micro-scale MTJ device via the tunnel magnetoresistance (TMR) effect. Our findings suggest that ultrafast spin transport, dipolar interactions, or a combination of both may contribute to the switching process, although the precise mechanism remains to be clarified. This work represents a significant step toward integrating ultrafast optical control with MTJ technology.
\end{abstract}

\maketitle

\section{\label{sec:level1}Introduction}
The ultrafast manipulation of magnetization has attracted considerable attention for both fundamental research and technological applications. The field of ultrafast magnetism was initiated by the discovery of ultrafast demagnetization—an abrupt reduction in magnetization within a picosecond—induced by femtosecond laser excitation~\cite{beaurepaire1996ultrafast}. Subsequently, in Co/Pt multilayer-based spin valves with symmetric ferromagnetic layers, it was observed that the demagnetization time depends on the initial magnetic configuration, either parallel (P) or antiparallel (AP) state~\cite{malinowski2008control}. This behavior has been attributed to nonlocal spin transport between the two ferromagnetic layers, triggered by ultrafast demagnetization in each layer~\cite{schellekens2014ultrafast,choi2014spin}. In particular, in the AP state, demagnetization is accelerated by spin currents carrying opposite spin polarization injected from the adjacent ferromagnetic layer. While such ultrafast spin currents can effectively manipulate magnetization dynamics, at that stage they did not lead to full magnetization reversal.

For many years, it was widely believed that single-shot all-optical switching (AOS) could not be achieved in purely ferromagnetic systems, in contrast to certain ferrimagnets such as Gd-based alloys, multilayers, and MnRuGa~\cite{radu2011transient,ostler2012ultrafast,lalieu2017deterministic,banerjee2020single,davies2020exchange,banerjee2021ultrafast}. In this context, Igarashi \textit{et al.} reported in 2023 that single-shot AOS can be realized in a Co/Pt-based spin valve by introducing structural asymmetry between the two ferromagnetic layers, such as differences in Co layer thickness and the number of Co/Pt repetitions~\cite{igarashi2023optically}. Depending on the laser fluence, two types of magnetization switching were observed: AP-to-P switching at high fluence and P-to-AP switching at low fluence. The former can be explained by nonlocal spin transport, as discussed above. The latter, however, is counterintuitive. Based on time-resolved experiments reported in a previous study~\cite{igarashi2023optically}, nonlocal spin transport is also believed to contribute to P-to-AP switching. To date, two possible scenarios have been proposed: (i) spin currents generated during the remagnetization of the reference layer, and (ii) spin currents produced by the demagnetization of the free layer, followed by spin-flip reflection at the interface between the Cu spacer and the reference layer~\cite{remy2023ultrafast}. Recently, detailed time-resolved experiments revealed that P-to-AP switching occurs on the timescale of the reference layer’s remagnetization, supporting the first scenario \cite{singh2025ultrafast}. Furthermore, it has been demonstrated that single-shot AOS can occur even without direct laser excitation, suggesting that complete demagnetization of the free layer in the spin valve—triggered by a rapid rise in electronic temperature due to ultrafast stimuli such as optical or thermal pulses—is a key requirement for AOS in such systems \cite{ishibashi2025single}.

In recent years, increasing attention has been paid to integrating magnetic tunnel junctions (MTJs)—the core components of spin-transfer torque magnetic random-access memory (STT-MRAM)—with AOS technology~\cite{chen2017all,kimel2019writing,aviles2020single,wang2022picosecond,mondal2023single,salomoni2023field}. In the earliest report, GdFeCo, a well-established all-optically switchable material, was used as the free layer in an MTJ structure. Although single-shot AOS could be detected via the tunnel magnetoresistance (TMR) effect~\cite{julliere1975tunneling}, the observed TMR ratio was relatively low, around 0.6\%~\cite{chen2017all}. Later studies introduced composite free layers composed of CoFeB exchange-coupled with all-optically switchable materials such as Co/Tb multilayers~\cite{aviles2020single,mondal2023single,salomoni2023field}, Gd/Co bilayers~\cite{wang2022picosecond}, and GdCo alloys with in-plane magnetic anisotropy \cite{geiskopf2025single} to improve the TMR ratio. However, the incorporation of rare-earth elements unavoidably results in lower TMR ratios than those achieved in standard CoFeB/MgO MTJs when annealed at temperatures above 300$^\circ$C, which are required for practical device fabrication.

Inspired by previous studies showing that demagnetization dynamics in symmetric MTJ structures depend on the initial magnetic configuration~\cite{he2013ultrafast}, similar to ferromagnetic spin valves~\cite{malinowski2008control}, we explore the possibility of single-shot laser-induced magnetization reversal in MTJs with asymmetric structures. These findings suggest that structural asymmetry could enable single-shot laser-induced magnetization reversal, as demonstrated in spin valves~\cite{igarashi2023optically}. By adjusting the MTJ capping layer thickness, we achieve single-shot laser-induced magnetization reversal in rare‑earth‑free CoFeB/MgO‑based MTJs without an external magnetic field, and detect the switching via the TMR effect in microscale devices.

\section{Experimental method}
In this section, we describe the MTJ stack, its characterization, and the experimental procedure for the single-shot laser-induced magnetization reversal measurements.

\subsection{MTJ stack}
\begin{figure}
    \centering
    \includegraphics[width=0.8\linewidth, trim=0mm 70mm 0mm 0mm, clip]{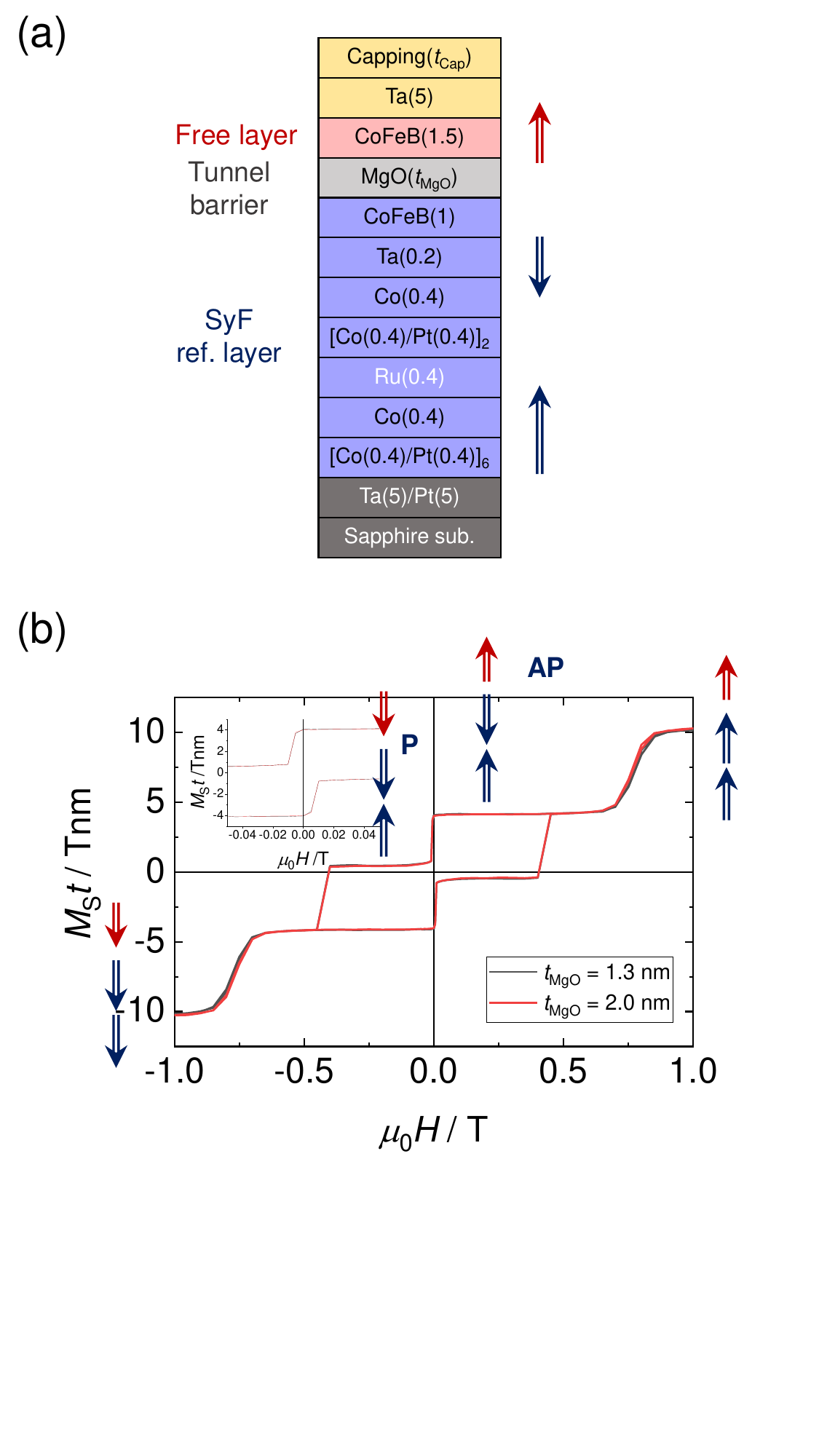}
    \caption{MTJ stack and its characterization. (a) Schematic of the MTJ stack structure used in this study. (b) $M$–$H$ curves measured for MTJ stacks with $t_\text{MgO} = 1.3$ and 2.0~nm. The inset shows a magnified view. Arrows indicate the corresponding magnetic configurations.}
    \label{fig1}
\end{figure}

We employ a de facto standard MTJ structure, consisting of a CoFeB free layer with an easy axis perpendicular to the sample plane~\cite{ikeda2010perpendicular}, and a synthetic ferrimagnetic (SyF) reference layer~\cite{sato2013mgo}, as illustrated in Fig.~\ref{fig1}(a). All samples were fabricated by sputtering. The basic layer stack is as follows (thicknesses in nanometers):  
Ta(5)/Pt(5)/[Co(0.4)/Pt(0.6)]$_{6}$/Co(0.4)/\allowbreak Ru(0.4)/[Co(0.4)/Pt(0.4)]$_{2}$/Co(0.4)/Ta(0.2)/CoFeB(1)/\allowbreak MgO($t_\text{MgO}$)/CoFeB(1.5)/Ta(5)/Capping($t_\text{cap}$),  
deposited on a sapphire or thermally oxidized Si substrate. The sample deposited on a sapphire substrate was used for magnetization reversal measurements, while the one deposited on a thermally oxidized Si substrate was used for sample characterization, as described in the next section. After deposition, the MTJ stack was annealed at 300$^\circ$C for 1~hour. We define the layers separated by the MgO barrier as the SyF reference layer (bottom, closer to the substrate) and the CoFeB free layer (top, farther from the substrate), as shown in Fig.~\ref{fig1}(a). The MgO layer thicknesses, $t_\text{MgO}$, are 1.3 and 2.0 nm. For the capping layer, we use either Ru or Pt. The capping layer thicknesses $t_\text{cap}$ used in this study are 0, 1.0, 2.0, 3.0, 4.0, and 5.0~nm for Ru, and 2.5 and 5.0~nm for Pt.

\subsection{Sample characterization}
Prior to the experiments, the MTJ stacks were characterized. Due to experimental constraints, the MTJ stacks were deposited on thermally oxidized Si substrates. It should be noted that the intrinsic stack properties are expected to be nearly identical regardless of the substrate used. In the present study, the MTJ stacks for characterization were deposited on thermally oxidized Si substrates, while the switching experiments were performed on sapphire substrates. The thermal conductivity of sapphire~\cite{cahill1998thermal} is higher than that of both SiO$_2$ and glass, while SiO$_2$ has a thermal conductivity comparable to that of glass~\cite{yamane2002measurement,van2000thermal}. Previous work has shown that the switching threshold is very similar for sapphire and glass substrates despite their large difference in thermal conductivity~\cite{igarashi2023optically}. This indicates that, in the ultrafast excitation regime relevant to the present study, the switching behavior is not strongly governed by substrate thermal conductivity. Therefore, the use of Si/SiO$_2$ and sapphire substrates is not expected to significantly affect the observed magnetization reversal, particularly if the underlying mechanism is governed by heat-driven processes, as reported in ferromagnetic spin valves~\cite{igarashi2023optically, ishibashi2025single, singh2025ultrafast}. Figure~\ref{fig1}(b) shows the magnetization curves measured using a vibrating sample magnetometer (VSM). A perpendicular easy axis is confirmed in both the free and reference layers. The corresponding magnetic configurations are illustrated in Fig.~\ref{fig1}(b). In this study, we focus on the P and AP states, as indicated in the figure.

Next, we evaluate the resistance-area product ($RA$) and the tunnel magnetoresistance (TMR) ratio using the current-in-plane tunneling (CIPT) method, which enables characterization of these parameters without device fabrication~\cite{worledge2003magnetoresistance}. Due to probe limitations, the measurement was performed only for the MTJ stack with $t_\text{MgO} = 1.3$~nm. The measured TMR ratio and $RA$ value are 158\% and 30~$\Omega\,\mu\mathrm{m}^2$, respectively. For $t_{\mathrm{MgO}} = 2.0$ nm, although a precise $RA$ value cannot be determined, an order-of-magnitude estimate can be obtained from device measurements, corresponding to an effective $RA$ on the order of $7 \times 10^{4}\ \Omega\,\mu\mathrm{m}^2$.

\subsection{Single-shot laser-induced magnetization reversal experiment}
The experimental setup used in this study is identical to that described in Ref.~\cite{igarashi2023optically}. Linearly polarized laser pulses were generated using a Ti:sapphire femtosecond laser system with a central wavelength of 800~nm and a repetition rate of 5~kHz. The pulse duration was varied from 50~fs to 10~ps. The 1/$e$ spot size $2\omega_0$ is approximately 80~$\mu$m.

To visualize the magnetic domain states, magneto-optical Kerr effect (MOKE) imaging was performed using a light-emitting diode (LED) with a center wavelength of 628~nm~\cite{iihama2018single}. The pump beam was incident on the sample at an angle of 45$^\circ$~\cite{igarashi2023optically}. A permanent magnet was used to apply an external magnetic field and to saturate the MTJ stack prior to laser pulse excitation. A schematic of the experimental setup is shown in Fig.~\ref{fig2}.

\section{Experimental results}
In this section, we show experimental results obtained with MTJ stacks.
\subsection{Single-shot laser-induced magnetization reversal experiment}
\begin{figure}
    \centering
    \includegraphics[width=0.8\linewidth, trim=0mm 220mm 0mm 0mm, clip]{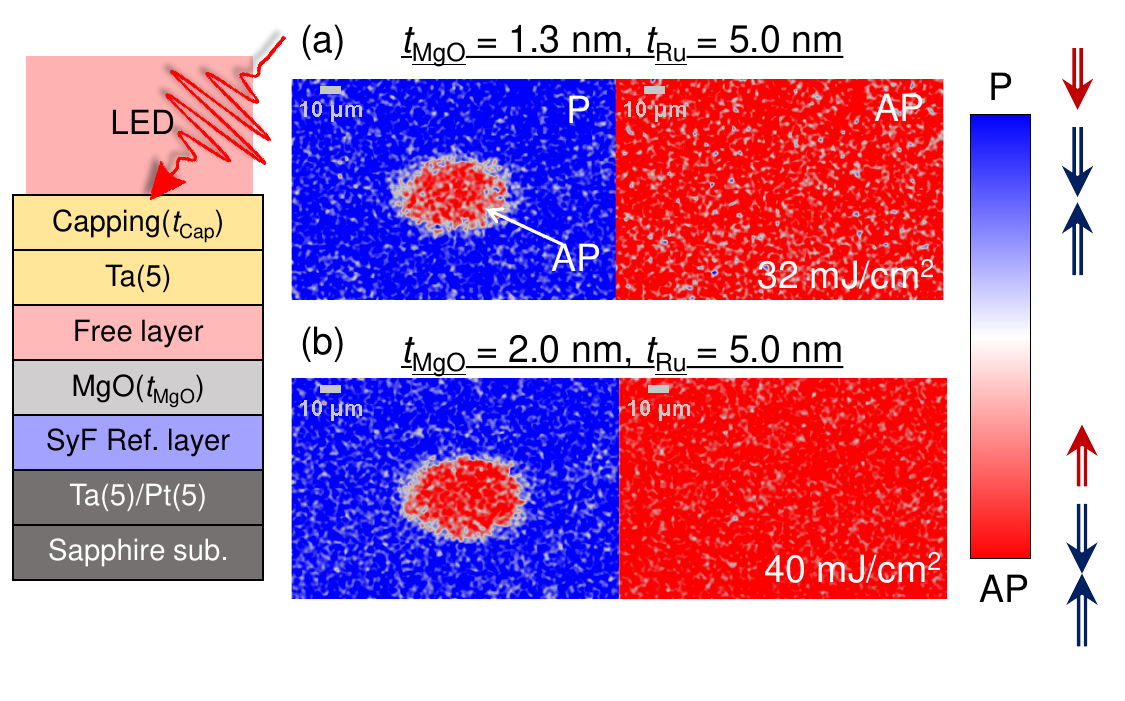}
    \caption{Single-shot magnetization switching observed via MOKE imaging in MTJ stacks with a Ru capping layer thickness of $t_\mathrm{Ru} = 5.0$ nm. (a) $t_\mathrm{MgO} = 1.3$ nm, (b) $t_\mathrm{MgO} = 2.0$ nm. For both cases, MOKE images are shown after laser irradiation, starting from the P (blue) and AP (red) states.}

    \label{fig2}
\end{figure}

Figure~\ref{fig2}(a) shows MOKE images obtained after laser pulse irradiation with a pulse duration of 50~fs for samples with a Ru capping layer thickness ($t_\mathrm{Ru}$) of 5.0~nm and $t_\text{MgO} = 1.3$~nm, initially in both the P (blue) and AP (red) states. While no AP-to-P switching is observed, clear P-to-AP switching is observed. We confirm that P-to-AP switching also occurs in a sample with $t_\text{MgO} = 2.0$~nm as shown in Fig. \ref{fig2}(b).

\begin{figure}
    \centering
    \includegraphics[width=0.7\linewidth, trim=0mm 0mm 0mm 0mm, clip]{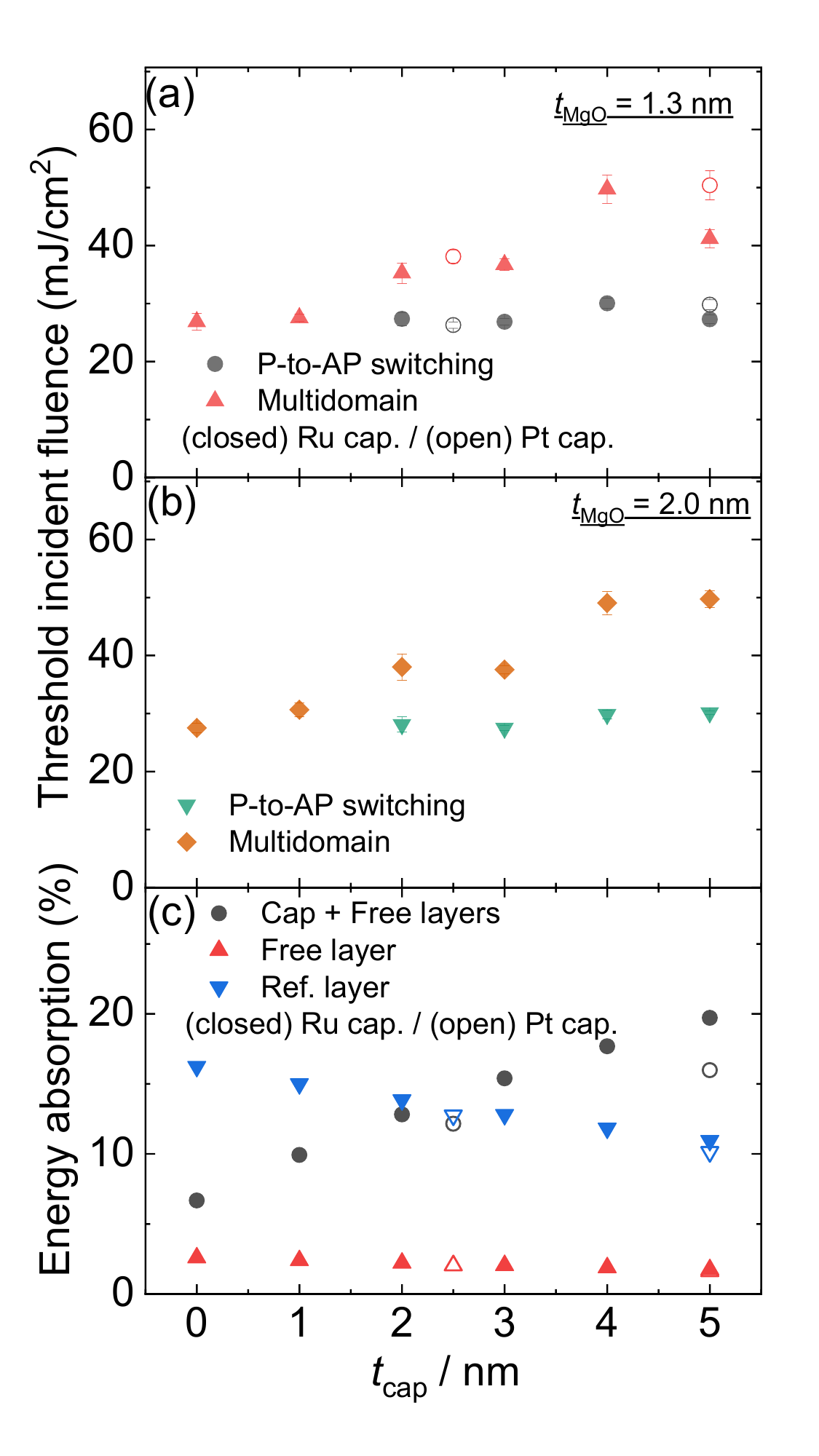}
    \caption{Summary of single-shot switching experiments in MTJ stacks. (a,~b) Evolution of the threshold incident fluences for P-to-AP switching ($F_{\text{P}}$) and multidomain formation ($F_{\text{MD}}$) as a function of capping layer thickness $t_{\mathrm{cap}}$ for (a) $t_\mathrm{MgO} = 1.3$\,nm and (b) $t_\mathrm{MgO} = 2.0$\,nm. (c) Calculated laser energy absorption in the capping, free, and reference layers as a function of $t_{\mathrm{cap}}$. Closed (open) symbols represent MTJ stacks with Ru (Pt) capping layers. Error bars represent the standard error obtained from fitting the equation described in the Experimental results section to the experimentally measured domain radius (area) as a function of laser pulse energy.}
    \label{fig3}
\end{figure}

For quantitative comparison, we extract the threshold incident fluences for P-to-AP switching ($F_{\text{P}}$) and multidomain formation ($F_{\text{MD}}$) by fitting the domain area as a function of laser pulse energy, as described in the Supplementary Material of Ref.\cite{igarashi2024influence}. Figures \ref{fig3}(a) and (b) summarize the threshold fluences as a function of $t_\text{cap}$ for $t_\text{MgO} =$ 1.3 and 2.0 nm, respectively. As shown in Figs. \ref{fig3}(a) and \ref{fig3}(b), the trends are essentially independent of MgO thickness. For P-to-AP switching, $F_\text{P}$ remains nearly constant with varying capping layer thickness, whereas $F_\text{MD}$ increases with $t_\text{cap}$. Notably, P-to-AP switching is observed only when $t_{\text{Ru}} \geq 2.0$ nm. Unlike previous studies on Co/Pt-based spin valves\cite{igarashi2023optically, igarashi2024influence}, AP-to-P switching is not observed in these MTJ stacks.

Since the only structural difference between the samples is the Ru capping thickness, we infer that laser-induced heating of the ferromagnetic layers in MTJ stacks plays a key role, similar to findings reported in Ref.\cite{ishibashi2025single}. To test this hypothesis, we performed the same experiment using Pt capping, which has comparable refractive indices to Ru. The results obtained with Pt capping are shown as open symbols in Fig. \ref{fig3}(a). As expected, $F_\text{P}$ with Pt capping is nearly identical to that with Ru capping. The $F_\text{MD}$ values with Pt capping also follow the same trend as those with Ru.

Furthermore, we calculate the laser energy absorption in the free and reference layers using the transfer matrix method, as shown in Fig. \ref{fig3}(c). The refractive indices of each material were obtained from Refs.~\cite{johnson1974optical, igarashi2020engineering}. As shown in Fig.~\ref{fig3}(c), increasing the thickness of the capping layer leads to a reduction in laser energy absorption in the reference layer. Although the energy absorption in the free layer also decreases with increasing capping thickness, the change is relatively small compared to that in the reference layer, because thicker Ru or Pt capping layers absorb a larger fraction of the incident laser energy. Consequently, the total energy absorption in the capping plus free layers increases monotonically with capping layer thickness, as indicated in the figure. In the present MTJ geometry, the calculation shows that only about 2.5\% of the incident laser energy is absorbed directly in the free layer (Fig. \ref{fig3}(c)). When the $F_\text{P}$ (27 mJ/cm$^2$) is renormalized by the energy actually absorbed in the free layer, the corresponding absorbed fluence is estimated to be approximately 0.675 mJ/cm$^2$, which is of the same order of magnitude as that reported previously for spin‑valve structures (0.25 mJ/cm$^2$)~\cite{igarashi2023optically}. These experimental and simulation results suggest that enhancing the heating of the free layer (including the capping layer), while suppressing laser‑induced heating of the reference layer, facilitates the observation of P‑to‑AP switching. This interpretation is consistent with conclusions drawn from previous studies using spin‑valve structures~\cite{igarashi2023optically, ishibashi2025single}. In addition to direct laser‑induced heating, ultrafast hot‑electron pulses generated by laser excitation of non‑magnetic layers have been shown to efficiently transport energy and induce ultrafast demagnetization~\cite{bergeard2016hot,bergeard2020tailoring,pudell2020heat}, and even magnetization reversal~\cite{wilson2017ultrafast,xu2017ultrafast,ishibashi2025single}. Such hot‑electron–mediated energy transport may also contribute to efficient heating of the free layer in the present MTJ system.

We also investigate pulse-duration dependence of the P-to-AP switching. See Appendix A for more details.
 
\subsection{Single-shot magnetization reversal detected via TMR effect}
\begin{figure}
    \centering
    \includegraphics[width=0.9\linewidth, trim=0mm 130mm 0mm 0mm, clip]{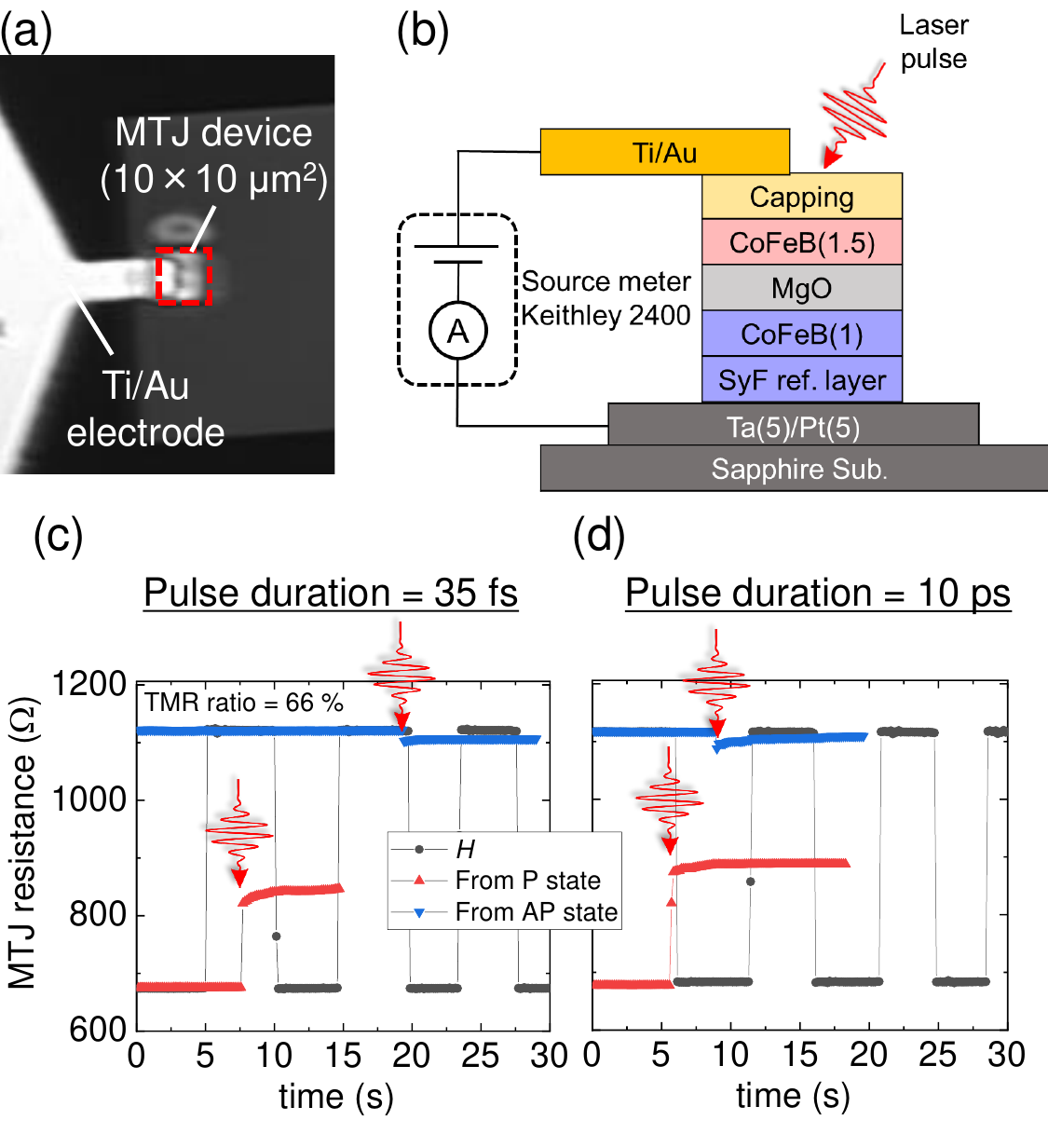}
    \caption{
    Single-shot switching detected via the TMR effect. (a) Optical microscopy image of the MTJ device. (b) Schematic of the experimental setup. (c,d) Time dependence of MTJ resistance in response to a square-wave magnetic field (black lines) and to irradiation of the P (red line) and AP state (blue line) by single laser pulses with the timing indicated by the laser pulse arrows. The pulse durations and incident fluences were (c) 35~fs and 21.6~mJ/cm$^2$, and (d) 10~ps and 41.0~mJ/cm$^2$, respectively.
 }
    \label{fig5}
\end{figure}
Having observed P-to-AP switching in MTJ stacks, our next objective is to detect P-to-AP switching via the tunnel magnetoresistance (TMR) effect in an MTJ device. To this end, we fabricated microscale MTJ devices with an MgO thickness of $t_\mathrm{MgO} = 2.0$ nm and $t_\mathrm{Ru} = 3.0$ nm. The fabrication process followed the procedure described in Ref.\cite{geiskopf2025single}. In this study, the ion milling was stopped at the Pt buffer layer to prevent domain nucleation in the reference layer caused by laser irradiation. An optical micrograph of the fabricated MTJ device is shown in Fig. \ref{fig5}(a). The device has lateral dimensions of $10 \times 10 \mu\mathrm{m}^2$. Due to technical constraints, the use of transparent electrodes was not feasible. Therefore, to enable optical access while simultaneously measuring the MTJ resistance, a Ti/Au electrode was designed to cover approximately 30 \% of the MTJ pillar, as in the previous study~\cite{geiskopf2025single}, as illustrated in Figs.~\ref{fig5}(a) and (b). 

We employ the same experimental setup as described in the previous study~\cite{geiskopf2025single}, in which a laser pulse is applied to the MTJ device while simultaneously measuring its resistance. The 1/$e$ spot size $2\omega_0$ is approximately 110~$\mu$m. A voltage of 30 mV is applied using a source meter (Keithley 2400) to monitor the MTJ resistance. Prior to laser irradiation, the MTJ device is initialized to either the P or AP state using a permanent magnet.

Figure~\ref{fig5}(c) shows the measured resistance of the MTJ device. The black symbols in Figs.~\ref{fig5}(c) and (d) represent magnetization reversal induced by an external magnetic field. After device fabrication, the TMR ratio is reduced from 158\% to 66\%. The reduction of the TMR ratio is attributed to processing‑induced effects, in particular metal redeposition along the sidewalls during ion milling, which can form parasitic conduction paths and effectively reduce the measured TMR, as previously reported~\cite{chen2003magnetic}. In addition, because the device resistance was measured in a two‑terminal configuration, series and contact resistances inevitably contribute to the measured resistance and further reduce the apparent TMR ratio.

Nevertheless, it is worth noting that the MTJ stack employed here is already used in STT‑MRAM production, indicating that such device‑level degradation can potentially be mitigated by optimized fabrication processes. Indeed, MTJ devices employing transparent electrodes have been demonstrated while maintaining high TMR ratios~\cite{olivier2020indium, shibata2024spin}. In Fig.~\ref{fig5}(c), a clear change in the MTJ resistance is observed only when the initial magnetic state is parallel (P) (red trace), indicating P‑to‑AP switching.

Importantly, electrode‑induced shadowing effects in MTJs with the same electrode configuration as in the present study have been quantitatively analyzed in previous studies using GdFeCo, a well‑established all‑optically switchable material~\cite{geiskopf2025single}. In particular, it has been shown that partial switching ratios and the associated resistance changes can be consistently explained by the opaque electrode geometry together with the Gaussian laser‑intensity profile. In the present device geometry, the Ti/Au top electrode covers approximately 30\% of the MTJ pillar, so that at most about 70\% of the junction area can be optically excited. As a result, the switched fraction is intrinsically limited by this geometric shadowing, and its exact value cannot be reliably extracted from the resistance signal alone. These considerations support the interpretation that the partial switching observed in the present MTJ devices originates from geometric shadowing effects rather than from an intrinsic limitation of the switching mechanism.

To further investigate the pulse-duration dependence, we also performed the same measurement using a 10 ps laser pulse. Consistent with the full-film pulse-duration data shown in Appendix A (Fig. \ref{fig4}), P-to-AP switching was also detected via the TMR effect in the patterned device under 10 ps excitation. The dependence of the switching behavior on laser fluence is summarized in Appendix B. 

\section{Discussion}
We have demonstrated single-shot P-to-AP switching in CoFeB/MgO-based MTJ stacks and devices, which adopt the same structure as those used in STT-MRAM technology. In this section, we discuss possible mechanisms underlying the observed P-to-AP switching.

While we initially expected that P-to-AP switching in MTJ stacks would result from ultrafast spin transport—as previously reported in ferromagnetic spin valves~\cite{igarashi2023optically}—the precise switching mechanism remains unresolved. In general, inserting an MgO barrier between conventional ferromagnetic layers strongly suppresses ultrafast spin-current transport compared to metallic spacers such as Cu. For MgO thicknesses exceeding 1~nm, spin transport is therefore expected to be significantly reduced in most heterostructures~\cite{wahada2022atomic, rouzegar2024terahertz}.

However, CoFeB/MgO junctions constitute a notable exception. In this system, electron tunneling is dominated by spin-polarized $\Delta_1$ symmetry states, leading to highly spin-selective transport across the MgO barrier and enabling very high TMR ratios exceeding 100\%~\cite{butler2001spin,mathon2001theory}. Indeed, previous work has reported signatures consistent with spin-current tunneling in CoFeB/MgO/CoFeB systems~\cite{he2013ultrafast}. Furthermore, previous work demonstrated that current‑induced STT switching can still be observed even in structures employing a multilayer free layer of [CoFeB(2.0)/MgO(0.93)]$_{4}$~\cite{igarashi2024single}. In the present work, the observation of a finite TMR even at $t_{\mathrm{MgO}} = 2.0$ nm confirms that electron tunneling across the MgO barrier remains active. While this does not by itself establish the presence of an ultrafast spin-current channel, it is consistent with a scenario in which spin-polarized transport may still play a role in the switching dynamics.

In addition to ultrafast spin transport, coupling between the free and reference layers may also contribute to the observed switching, analogous to mechanisms discussed in the context of heat-assisted magnetic recording (HAMR)~\cite{kryder2008heat}. To explore this possibility, we measured the magnetic hysteresis loops using the magneto-optical Kerr effect (MOKE), as presented in Appendix~C. Figure~\ref{fig7} reveals a small shift field that biases the MTJ toward the AP state. The extracted bias-field values are approximately $-1.05$~mT and $+1.15$~mT for the two magnetic configurations. Such a bias field may influence the final magnetic state during strong laser-induced demagnetization; however, it is not obvious that a static field of this magnitude alone is sufficient to deterministically select the AP state under ultrafast excitation. Moreover, the temporal evolution of the effective bias field during laser excitation—when both the free and reference layers are driven far from equilibrium—is itself not trivial and cannot be inferred from static measurements. At the same time, the nearly identical P‑to‑AP switching thresholds observed for $t_{\mathrm{MgO}} = 1.3$ and $2.0$~nm do not support a switching mechanism dominated purely by spin‑current transport through the MgO barrier. Taken together, the present results indicate that neither a purely ultrafast spin-current-driven picture nor a purely HAMR-like scenario can be uniquely identified based on the current dataset.

Time-resolved MOKE (TR-MOKE) experiments could provide further insights by distinguishing between the two possible mechanisms—ultrafast spin-current-induced switching and thermally assisted field-driven switching—due to their different timescales~\cite{igarashi2024influence}. However, the complex multilayer structure of the MTJ stacks used in this study makes TR‑MOKE measurements technically challenging. The free layer is only 1.5 nm thick, whereas the reference layer has an effective thickness exceeding 10 nm. As a result, the Kerr signal from the free layer is extremely weak and largely masked by contributions from the reference layer and the surrounding multilayer stack.
We did attempt TR‑MOKE measurements in a reflection‑geometry configuration; however, the detected signal was dominated by the reference layer, and no reliable free‑layer‑resolved dynamics could be extracted with a sufficient signal‑to‑noise ratio. Under these conditions, quantitative tracking of the free‑layer magnetization dynamics was not feasible. If similar P-to-AP switching can be achieved in a simplified MTJ stack, for example, one without a SyF reference layer, it may be possible to directly measure the magnetization reversal dynamics using TR-MOKE techniques.

Although the detailed mechanism of the switching remains to be clarified, our results demonstrate that single-shot magnetization reversal can be induced by carefully controlling the energy absorption ratio between the free and reference layers. In this proof-of-concept study, we employed a free layer consisting of a single CoFeB layer. However, the proposed approach is applicable to more practical structures, such as [CoFeB/MgO] multilayers, which are better suited for memory applications \cite{sato2012perpendicular, nishioka2019novel, igarashi2024single}. Importantly, the energy-absorption-profile strategy demonstrated in this study is not limited to CoFeB/MgO MTJs, but can in principle be extended to other magnetic heterostructures, such as FePt-based spin-valve systems~\cite{tozman2024dual}. 

\section{Conclusion}
We have demonstrated single-shot laser-pulse-induced magnetization reversal in CoFeB/MgO-based MTJs without relying on rare-earth elements, using a structure compatible with STT-MRAM technology. Although complete switching of the entire MTJ area was not achieved due to the shadowing effect of the electrode~\cite{geiskopf2025single}, we have shown that this single-shot magnetization reversal can be electrically detected via the TMR effect in a microscale MTJ device.

Our combined experimental and numerical results demonstrate that modifying the laser energy absorption profile within the MTJ stack is a key factor in enabling magnetization reversal. While the underlying physical mechanisms remain to be fully understood, this study establishes a crucial link between ultrafast magnetism and spintronics device platforms. Our findings represent an important step toward the integration of ultrafast optical control into practical MTJ-based memory technologies.

\begin{acknowledgments}
We thank E. D\'{i}az, A. Anad\'{o}n, J.-X. Lin, L. D. Buda-Prejbeanu, R. C. Sousa, I. L. Prejbeanu, K. Ishibashi, S. Iihama, J.-Y. Yoon, T. Arakawa, N.-H. Kaneko for fruitful discussions. This work was supported by the French National Research Agency (ANR) through the France 2030 government PEPR Electronic grants EMCOM (ANR-22-PEEL-0009) by the ANR SPOTZ project ANR-20-CE24-0003, SLAM ANR- 23-CE30-0047, the project MAT-PULSE from “Lorraine Universit\'{e} d’Excellence” reference ANR-15-IDEX-04-LUE, the Institute Carnot ICEEL, the R\'{e}gion Grand Est, the M\'{e}tropole Grand Nancy, the “FEDERFSE Lorraine et Massif Vosges 2014-2020”, a European Union Program,  the Academy of Finland (Grant No. 316857), the Sakura Program, the JSPS Bilateral Program, JSPS KAKENHI JP 24K22964, 24H02235, 24H00039, 26K01371, MEXT Initiative to Establish Next-generation Novel Integrated Circuits Centers (X-NICS) (Grant No. JPJ011438), the Tohoku University-Universit\'{e}  de Lorraine Matching Funds, and CSIS cooperative research project in Tohoku University. The devices in the present study were patterned at MiNaLor platform which is partially funded by Grand Est Region and FEDER through RaNGE project. This article is based upon work from COST Action CA23136 CHIROMAG, supported by COST (European Cooperation in Science and Technology). J.I. acknowledges support from JSPS Overseas Research Fellowships.
\end{acknowledgments}

\section*{Author Declarations}
St\'{e}phane Mangin is a Guest Editor of APL Materials. To avoid any potential conflict of interest, he was not involved in the editorial handling or decision-making process for this manuscript.

\section*{Data Availability Statement}
The data that support the findings of this study are available from the corresponding author upon reasonable request.

\section*{References}
\nocite{*}
\bibliography{AOSMTJ}

\appendix

\section{Pulse duration dependence}

\begin{figure}
    \centering
    \includegraphics[width=0.8\linewidth, trim=0mm 100mm 0mm 0mm, clip]{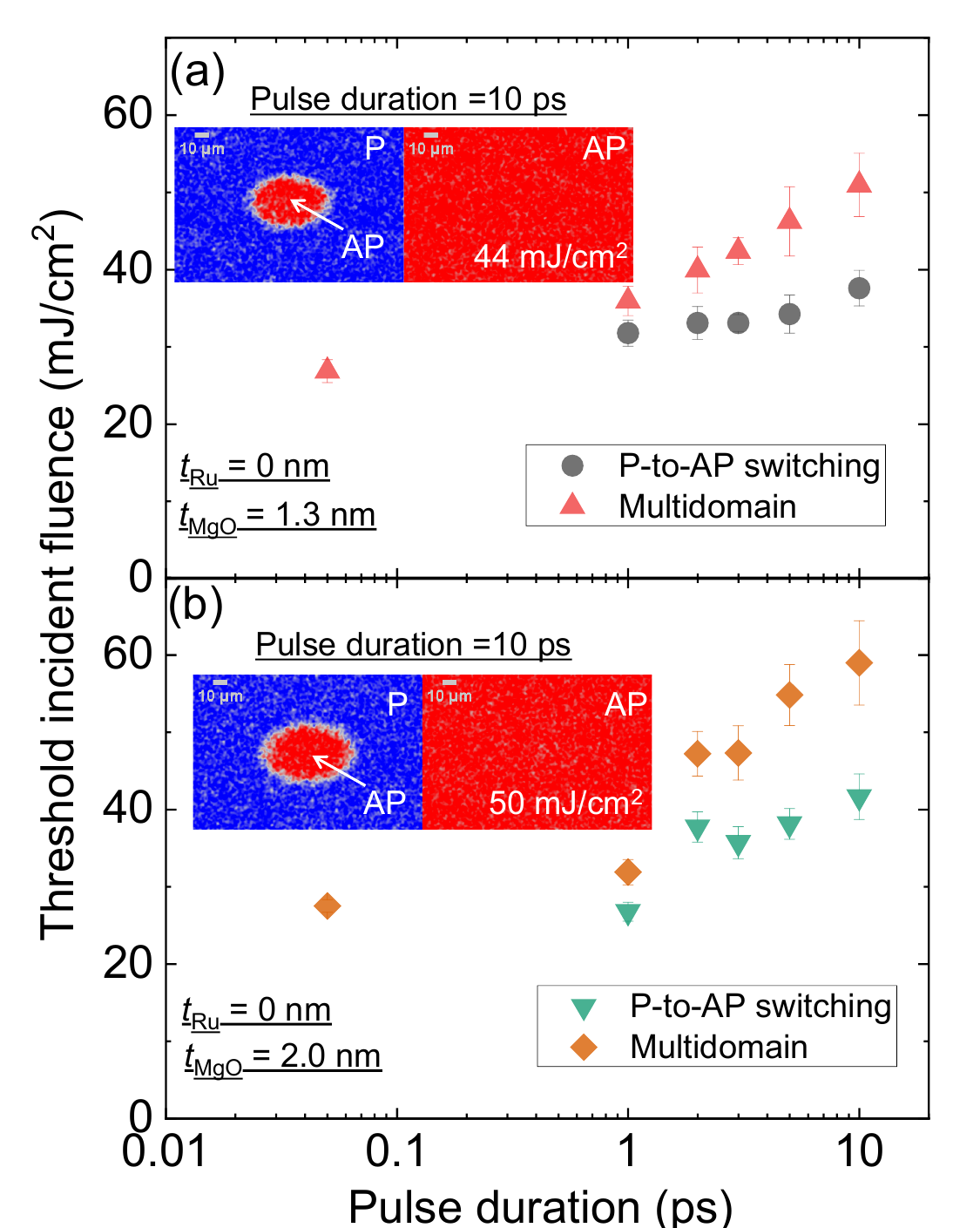}
    \caption{
    (a,~b) Evolution of the threshold incident fluences for P-to-AP switching ($F_{\text{P}}$) and multidomain formation ($F_{\text{MD}}$) as a function of pulse duration for 
    (a) $t_\mathrm{MgO} = 1.3$\,nm and 
    (b) $t_\mathrm{MgO} = 2.0$\,nm. Error bars represent the standard error obtained from fitting the equation described in the Experimental results section to the experimentally measured domain radius (area) as a function of laser pulse energy.
    }
    \label{fig4}
\end{figure}

Here we present the pulse duration dependence of P-to-AP switching and multidomain formation. Figure~\ref{fig4} summarizes the evolution of the threshold incident fluences for both P-to-AP switching and multidomain states as a function of pulse duration for samples with $t_\mathrm{Ru} = 0$~nm and $t_\mathrm{MgO} = 1.3$ and 2.0~nm. As shown in Fig.~\ref{fig3}, for a pulse duration of 50~fs, P-to-AP switching is observed only when $t_{\text{Ru}} \geq 2.0\,\mathrm{nm}$. However, even for $t_{\mathrm{Ru}} = 0$~nm, P-to-AP switching begins to emerge as the pulse duration increases in both structures. The threshold incident fluence for P-to-AP switching increases gradually with pulse duration, whereas that for multidomain formation increases more rapidly. This contrast in their pulse-duration dependence is consistent with trends reported in previous studies~\cite{igarashi2023optically, ishibashi2025single}. The underlying mechanism for the observed P-to-AP switching at a pulse duration of 10~ps remains unclear. A previous study on ferromagnetic spin valves showed that the inclusion of a Cu heat sink extends the switching window up to 8~ps\cite{ishibashi2025single}. In the present study, no such heat sink is applied to the MTJ stacks. One possible explanation is the use of sapphire substrates in this work, which offer superior heat dissipation compared to the glass substrates used in previous studies on ferromagnetic spin valves~\cite{igarashi2023optically}.

\section{Fluence dependence}
\begin{figure}
    \centering
    \includegraphics[width=1\linewidth, trim=0mm 250mm 0mm 0mm, clip]{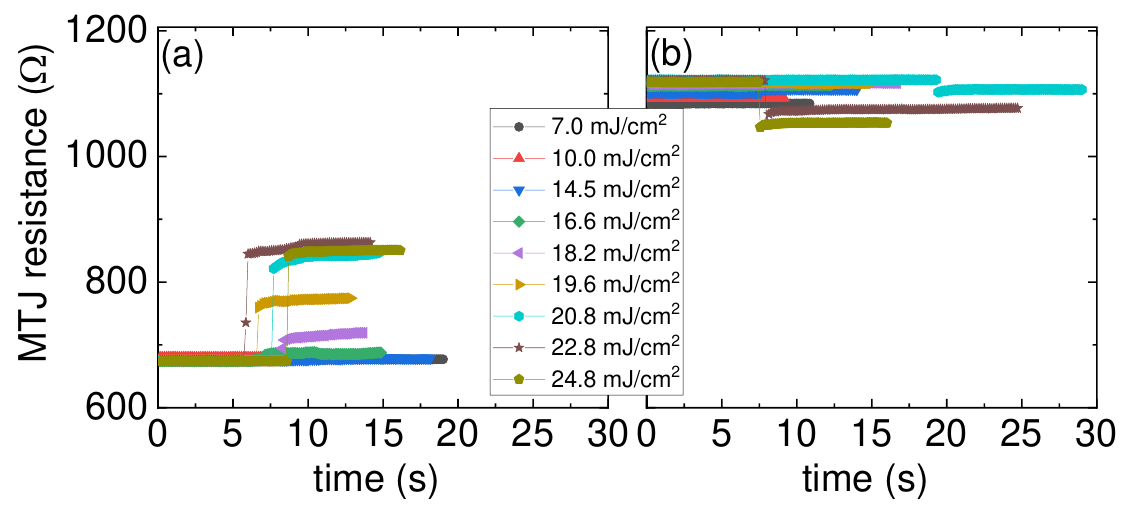}
    \caption{
    Time evolution of MTJ resistance under various incident fluences with a pulse duration of 35 fs. (a) Initial magnetization in the parallel (P) state. (b) Initial magnetization in the antiparallel (AP) state.
    }
    \label{fig6}
\end{figure}
Here we present the results obtained under various laser fluences. Figure~\ref{fig6} shows the time evolution of MTJ resistance for different incident fluences (pulse duration: 35~fs). When the initial state is parallel (P), the MTJ resistance starts to increase upon laser irradiation at an incident fluence of 18.2~mJ/cm$^2$, indicating P-to-AP switching. In contrast, when starting from the antiparallel (AP) state, no significant resistance change is observed at comparable fluences. However, at incident fluences exceeding 19.6~mJ/cm$^2$, a slight decrease in resistance is observed even when the initial state is AP, suggesting the onset of multi-domain formation in the MTJ device at higher incident fluences. 
Although the obtained threshold fluence may appear high, it is important to emphasize that it corresponds to the incident laser fluence. As shown in Fig.~3(c), only about 2.5\% of the incident energy is directly absorbed in the free layer. The total optical absorption of the MTJ stack is approximately 38\%, indicating that more than half of the incident laser energy is reflected.

The robustness of magnetic heterostructures under intense ultrafast excitation has been demonstrated in several previous studies. In particular, our recent work employed even higher incident fluences exceeding 100~mJ/cm$^2$ without any observable degradation of magnetic properties~\cite{ishibashi2025single}.

In the present study, the laser fluence was systematically varied, and the corresponding time evolution of the MTJ resistance is shown in Fig.~\ref{fig6}. All measurements were performed in a regime where the magnetic state could be reproducibly reset using a permanent magnet. The distinctly different responses observed for the initial P and AP states indicate that the resistance changes originate from the magnetic configuration rather than from laser-induced damage, thermally driven resistance drift, or MgO barrier breakdown.

\section{MOKE Hysteresis Measurements}
\begin{figure}
    \centering
    \includegraphics[width=0.8\linewidth, trim=0mm 90mm 0mm 0mm, clip]{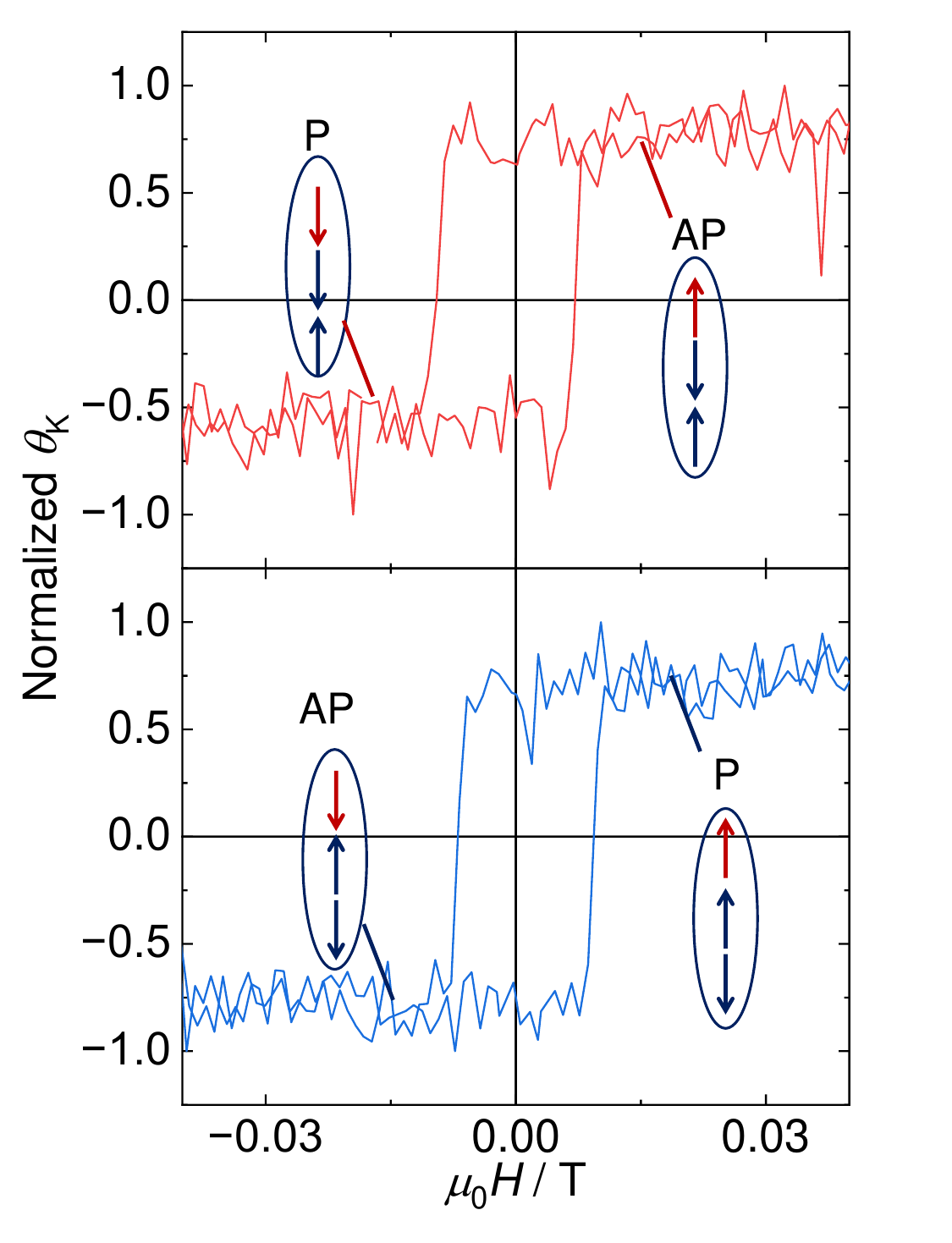}
    \caption{
    Minor MOKE hysteresis loops of the MTJ stack. Arrows indicate the magnetic configuration corresponding to each MTJ state (P or AP). The wine-red and navy arrows represent the magnetization directions of the free and reference layers, respectively.
    }
    \label{fig7}
\end{figure}
Here we present the results of MOKE hysteresis loop measurements performed on the MTJ stack. Figure~\ref{fig7} shows minor loops for two different magnetic configurations of the reference layer. The extracted bias-field values are approximately $-1.05$ mT and $+1.15$ mT for the two configurations. Providing a quantitative model for the time evolution of the bias field during laser-induced transient heating is challenging in the present system, particularly because the reference layer has a SyF structure. The instantaneous magnitude—and even the sign—of the effective bias field may depend on the balance of ultrafast magnetization dynamics within the SyF reference layer immediately after excitation. Addressing this rigorously would require time-resolved measurements such as TR-MOKE, which are technically difficult for the current multilayer MTJ stack.
Nevertheless, from a physical standpoint, it is reasonable to state that during laser excitation—when the magnetization amplitudes of the magnetic layers are reduced—the effective bias field is unlikely to exceed the values extracted from the static MOKE loops by more than an order of magnitude. The MOKE-derived bias field therefore serves as a realistic upper bound for the effective field acting on the strongly demagnetized free layer.

\end{document}